\documentclass[aps,prmaterials,10pt,twocolumn,nofootinbib,showpacs,longbibliography,floatfix]{revtex4-2}
\usepackage{graphicx} 
\usepackage{amssymb}
\usepackage{amsmath}
\usepackage{placeins} 
\usepackage{tabularx}
\usepackage{makecell}
\usepackage{calc}
\begin{document}
\newcommand{\aSn}{{$\alpha$-Sn}}
\newcommand{\aSnGe}{{$\alpha$-Sn$_{0.98}$Ge$_{0.02}$}}
\newcommand{\atn}{${\alpha}$-Sn}
\newcommand{\figintro}{
\begin{figure*}
    \centering
    a) \raisebox{1ex-\height}{\includegraphics[width=2in]{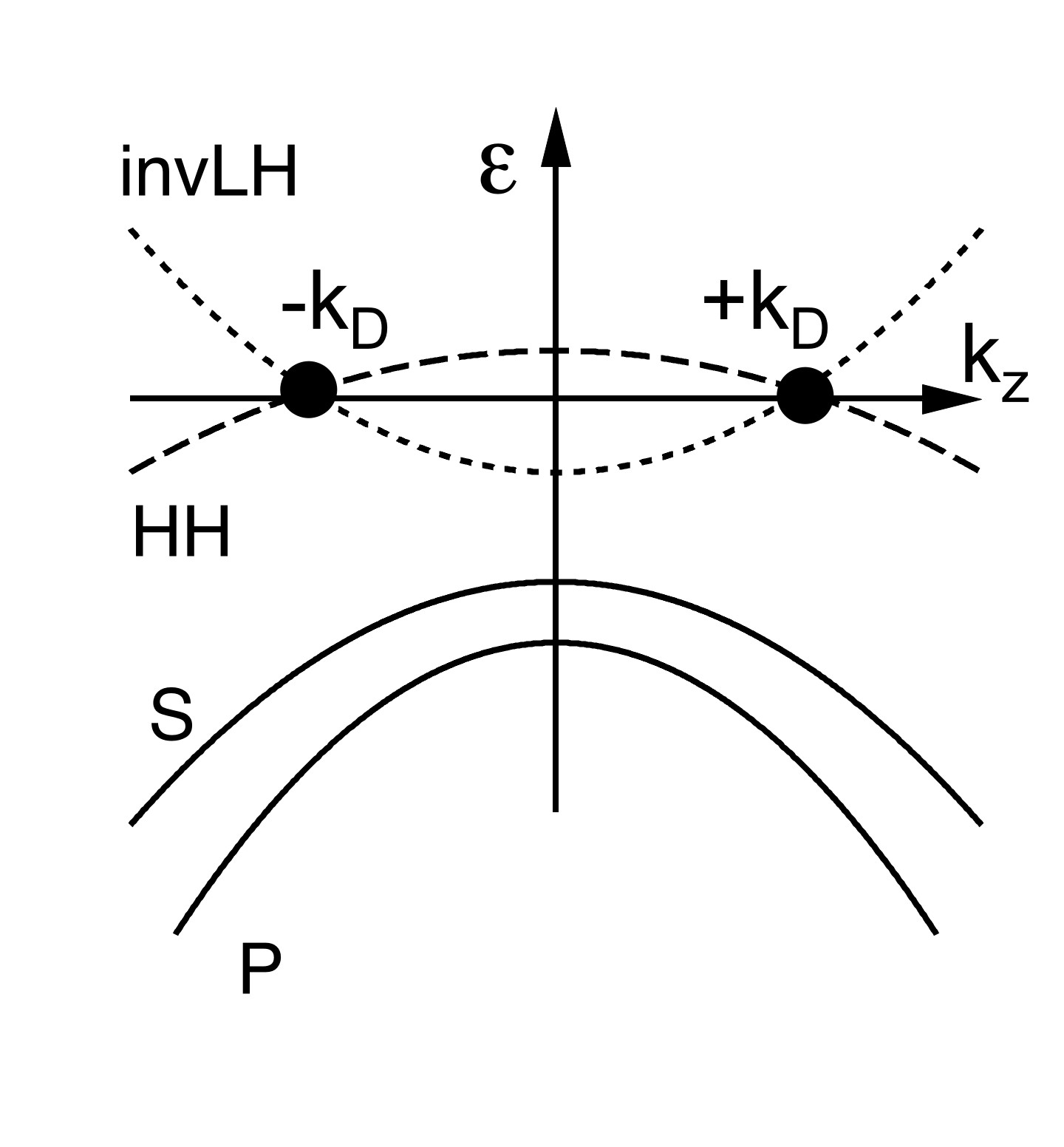}}
    ~~~~ b) \raisebox{1ex-\height}{ \includegraphics[width=3.5in]
      {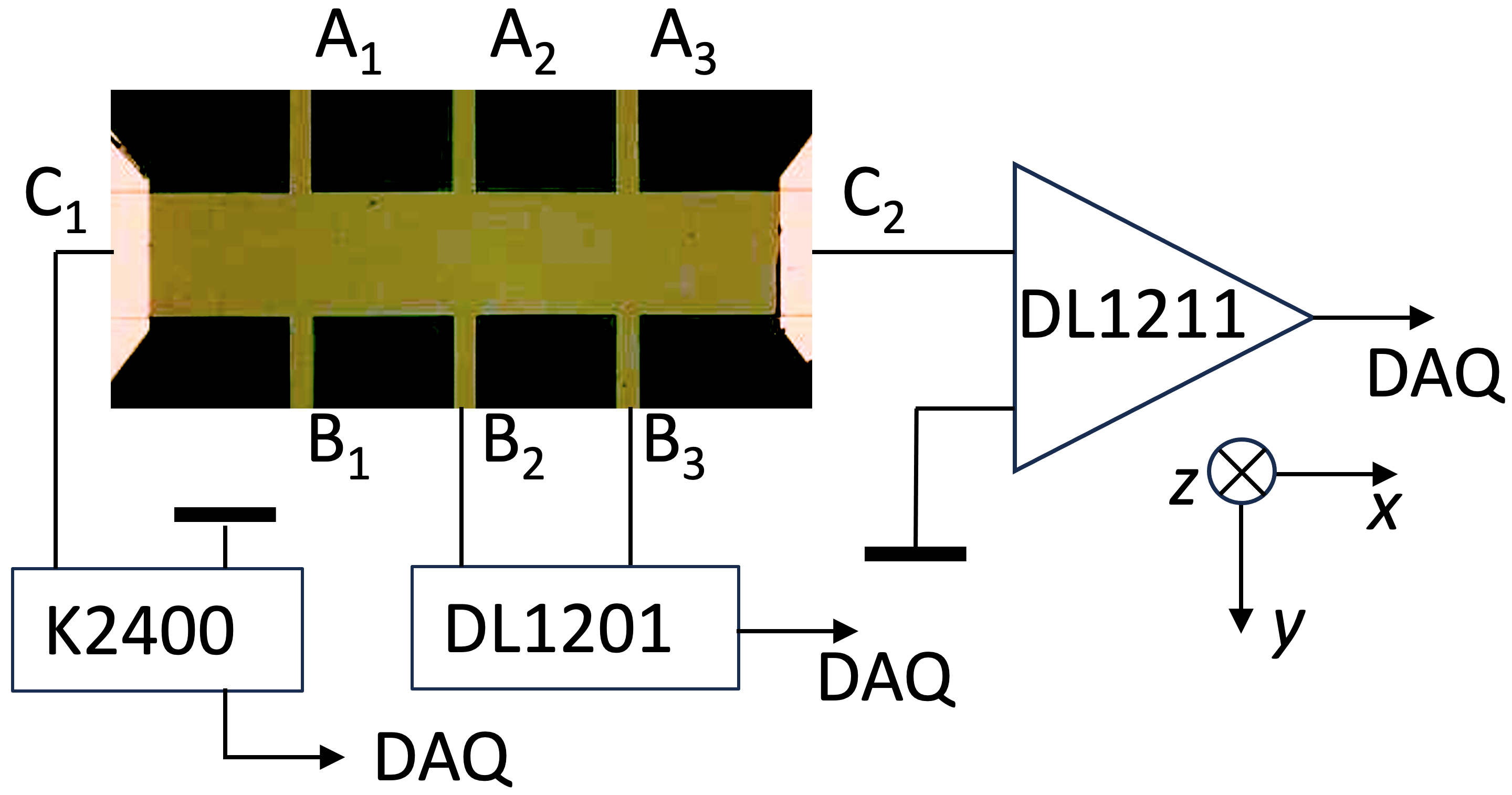}}  
    \caption{a) Dirac points $k=(0,0,\pm k_d)$ in a strained  \atn{}  film under a tensile strain normal to the film plane due  to  a ``negative gap" between the inverted light-holes band (invLH) and the heavy-holes band (HH). b) Schematic of the Hall bar and electronics configuration for the magnetotransport measurement.  The coordinate system defines directions relative to the Hall bar.}
    \label{figintro}
\end{figure*}
}

\newcommand{\figa}{
\begin{figure*}
    \centering
    \includegraphics[width=6.5in]{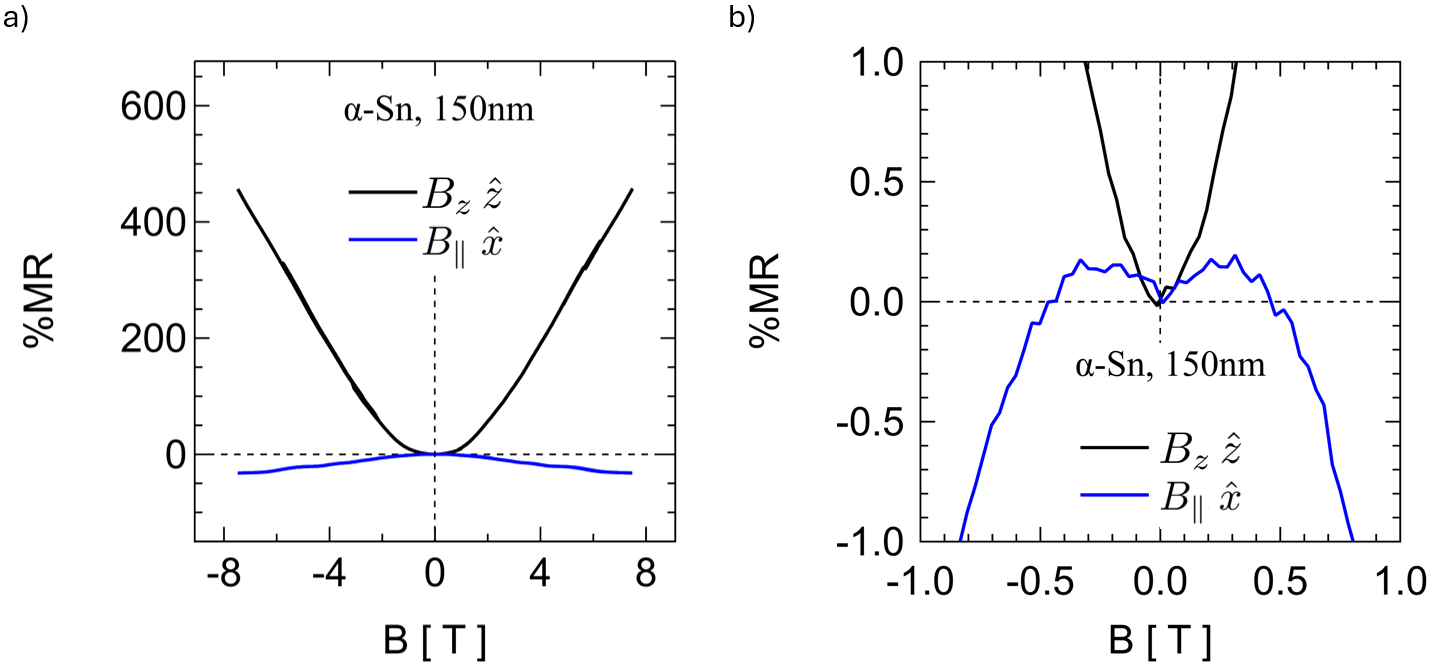}
    \caption{a) Magnetoresistance of a strained $\alpha-$Sn epilayer grown on CdTe with magnetic field $B_z$ normal to the film plane (black trace) and $B_{||}$ parallel to the current direction (blue trace), at T= 5K. b): the expanded view of the $|B|<1$ T data showing negative MR  when $B_{||}\ge 0.5$ T}.
    \label{figa}
\end{figure*}
}

\newcommand{\figcdcombo}{
\begin{figure*}
 \includegraphics[width=6in]{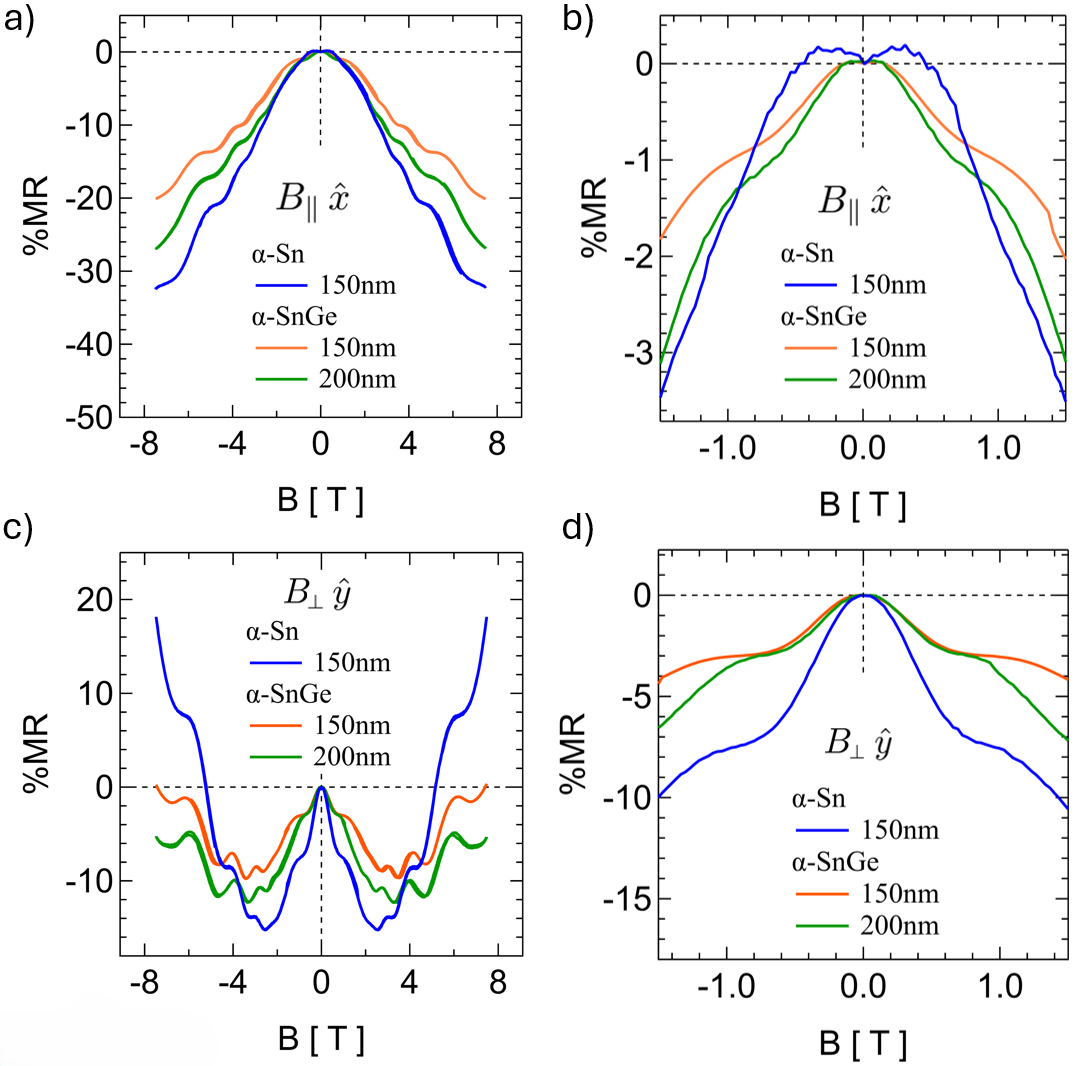}
\caption{Magnetoresistance MR with the magnetic field parallel to the film plane. a) $\vec{B} = B_\parallel \hat{x}$, monotonic negative MR up to maximum field b) $\vec{B} = B_\parallel \hat{x}$, expanded view of $|B|<1.5$ T  c) $\vec{B} = B_\perp \hat{y}$, negative MR up to $|B|<5.5$ T , then MR follows an increasing trend d) $\vec{B} = B_\perp \hat{y}$, expanded view of $|B|<1.5$ T. All samples show qualitatively the same MR behavior with in-plane field and Ge-alloying slightly reduces the field effects on the samples.}
 \label{figcdcombo}
\end{figure*}
}

\newcommand{\figxrd}{
\begin{figure}
    \centering
   
    \includegraphics[width=2.5in]{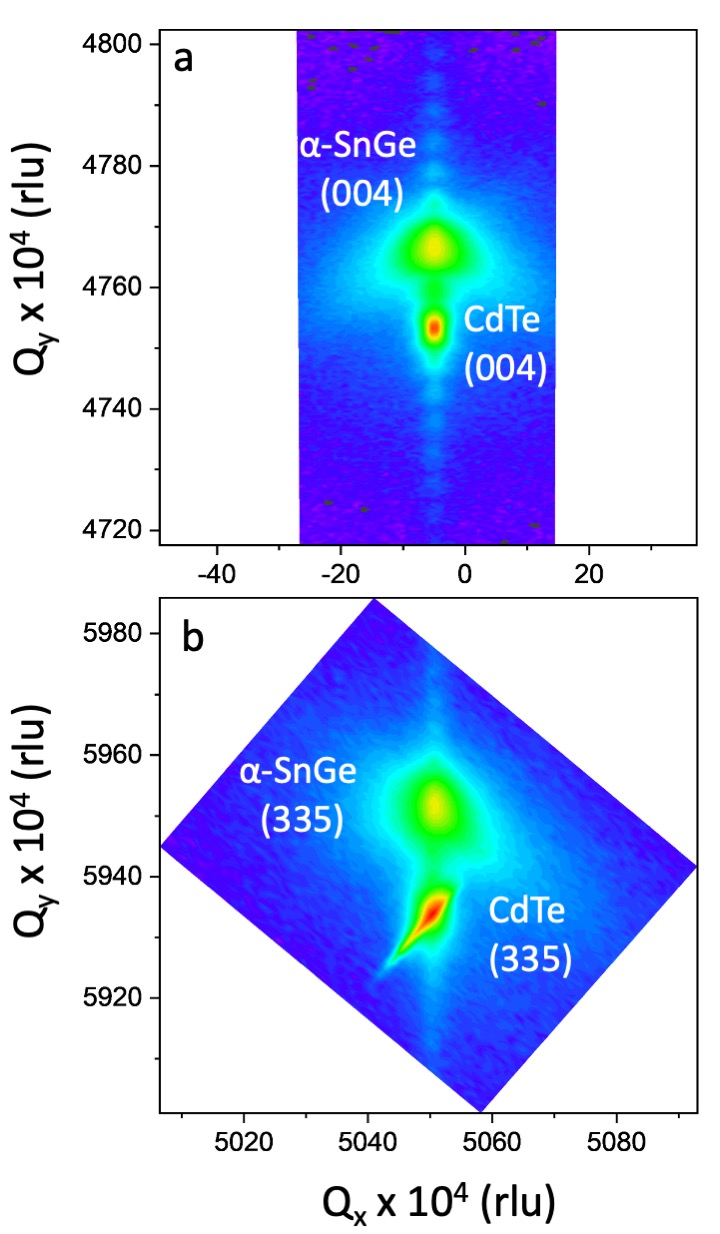}
    \caption{Reciprocal space maps of the a) (004), and b) (335) Bragg reflections for sample S2 plotted in dimensionless reciprocal lattice units (rlu) on a logarithmic intensity scale. Film and substrate peaks are annotated.}
    \label{figxrd}
\end{figure}
}

\newcommand{\figdevdim}{
\begin{figure}
    \centering
    \includegraphics[width=2.25 in]{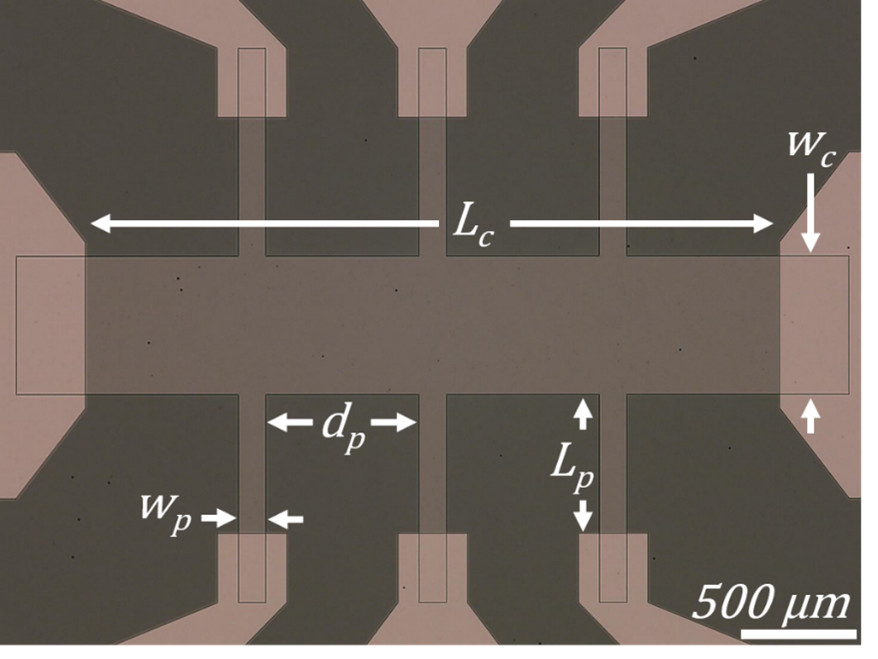}
    \caption{Optical micrograph of a finished Hall bar structure fabricated from an $\alpha$-Sn film on CdTe. Feature dimensions are in the main text.}
    \label{figdevdim}
\end{figure}
}

\newcommand{\figxrdd}{
\begin{figure}
    \centering
    \includegraphics[width=3.2in]{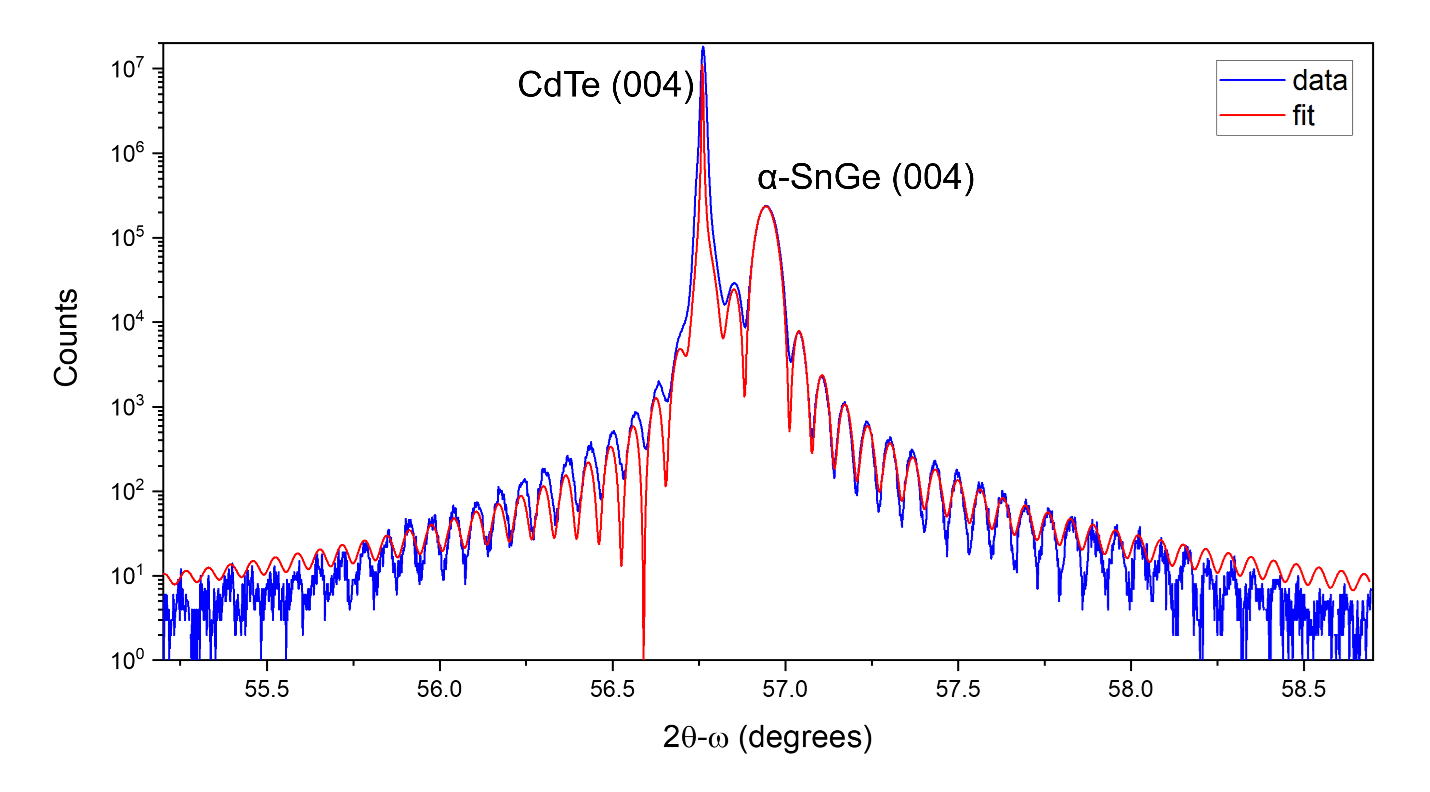}
    \caption{Out-of-plane XRD $2\theta - \omega$ scan data for sample S2. A fit using a one-layer model with homogeneous Sn and Ge concentration is overlaid.  Film and substrate peaks are annotated.  The relative peak positions indicate the film is under biaxial tensile strain.}
    \label{figxrdd}
\end{figure}
}

\newcommand{\fighallge}{
\begin{figure}
    \centering
    \includegraphics[width=3.2in]{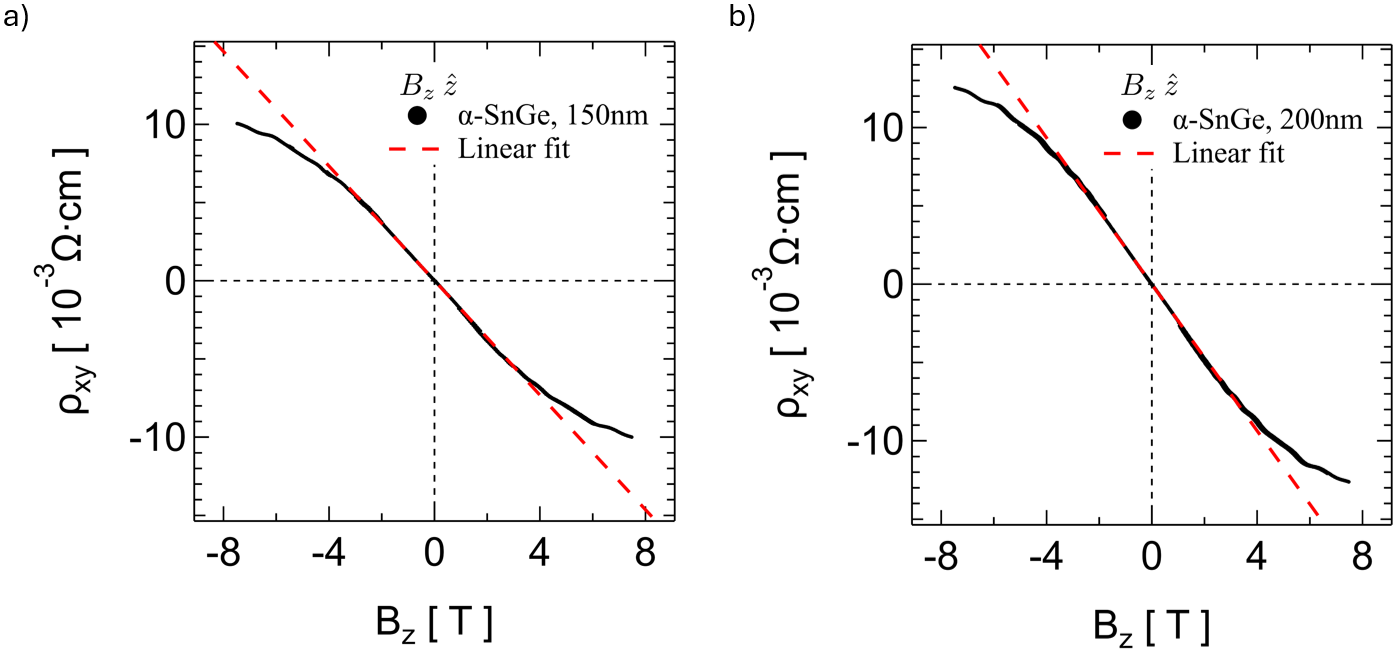}
    \caption{Black trace: measured Hall resistivity of Ge-alloyed \atn{} with magnetic field B normal to the film plane. Red trace: linear fit, up to B = 1 T. The negative slope corresponds to n-type carriers. a) Sample S2, b) Sample S3.}
    \label{LinearGeFit}
\end{figure}
}

\newcommand{\ttable}{
\begin{table*}
\centering
    \begin{tabular}{c|c|c|c|c|c|c|c|c|c|c|c}
        \toprule
        ~~&~~&Thickness&\multicolumn{2}{c|}{Substrate}&\multicolumn{7}{c}{Film}\\
        Sample& \% Ge & (nm) &a,b(\AA)&c(\AA)&a,b(\AA) &c(\AA) &$\varepsilon_{\parallel}$ & $\varepsilon_{\perp}$&relaxed (\AA)
        &\makecell{$\Omega/\square$ \\ ~~T=300 K ~~} &\makecell{$\Omega/\square$ \\ ~~T=5 K~~~ } \\
        \hline
        ~~&~~&~~&~~&~~&~~&~~&~~&~~&~~&~~&~~\\
        S1& 0.00& 149& 6.483& 6.483& 6.482& 6.496&-0.0010&0.0010& 6.490& 33.1&132\\

        S2& 1.90& 155& 6.483& 6.481& 6.482& 6.464&0.0013&-0.0013&6.473& 56.0&225\\

        S3& 1.86& 206& 6.483& 6.481& 6.482& 6.464&0.0013&-0.0013&6.473&38.2&179\\

    \end{tabular}
    \caption{Sample characteristics: germanium composition, film thickness, in-plane lattice parameters (a, b), out-of-plane lattice parameter (c), in-plane strain $\varepsilon_{||}$, out-of-plane strain $\varepsilon_\perp$, the lattice parameter of the film if fully relaxed.}
\label{table1}
\end{table*}
}
\newcommand{\tttable}{
\begin{table}
\begin{tabular}{c|c|c|c}
\hline \hline
~~ & ~~ & ~~ & ~~ \\
Sample ~& ~$\frac{d\rho_{xy}}{dB}$ ($10^{-3} \ \frac{\Omega \cdot cm}{T}$)~ & \makecell{~n ($10^{17} \ cm^{-3}$)} ~ & \makecell{~$\mu_{tr}$ ($\frac{cm^2}{V \cdot s}$)~} \\
~~ & ~~ & ~~ & ~~ \\
\hline
~~ & ~~ & ~~ & ~~ \\
S2 &$-1.831 \pm 0.004$&3.41&5424\\
S3 &$-2.338 \pm 0.002$&2.67&6532\\
\end{tabular}
\caption{4 K transport properties of $\alpha$-SnGe samples.}
\label{TransportTable}
\end{table}
}

\newcommand{\figmrbz}{
\begin{figure}
    \centering
    \includegraphics[width=2.5 in]{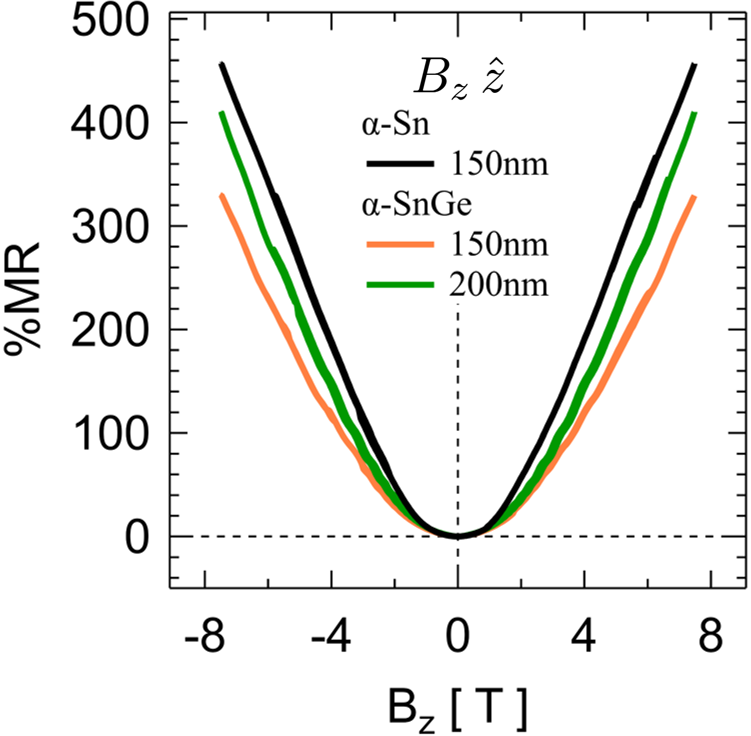}
    \caption{Magnetoresistance of strained \atn{} and $\alpha$-SnGe with $B_z$ normal to the films' plane. All samples shows qualitatively the same field dependence with MR $\approx$ 400\% at maximum field and Ge-alloying slightly reduces the field effect.}
    \label{mrbz}
\end{figure}
}
\title{Negative magnetoresistance in strained $\alpha$-Sn  and $\alpha$-SnGe films in an in-plane magnetic field}
\author{Sunny Phan}
\affiliation{Department of Physics and Astronomy, University of Cincinnati, Cincinnati, OH 45221, USA} 
\author{Jesse Thompson}
\affiliation{KBR, 3725 Pentagon Blvd Suite 210, Beavercreek Township, OH 45431, USA}
\affiliation{Air Force Research Laboratory, Wright-Patterson AFB, OH 45433, USA}
\author{Trent Johnson}
\affiliation{KBR, 3725 Pentagon Blvd Suite 210, Beavercreek Township, OH 45431, USA}
\affiliation{Air Force Research Laboratory, Wright-Patterson AFB, OH 45433, USA}
\author{Alexander Khaetskii}
\email{khaetskii@gmail.com}
\affiliation{Department of Physics and Astronomy, Ohio University, Athens, OH 45701, USA}
\author{Andrei Kogan}
\email{andrei.kogan@uc.edu}
\affiliation{Department of Physics and Astronomy, University of Cincinnati, Cincinnati, OH 45221, USA}  
\author{Arnold Kiefer}
\affiliation{Air Force Research Laboratory, Wright-Patterson AFB, OH 45433, USA}
\date{\today}
\begin{abstract}
To test the hypothesis that the chiral anomaly is responsible for negative magnetoresitance (MR) in \atn{}, we have studied magnetotransport in strained, epitaxial films of pure \aSn{} and the alloy \aSnGe{} that are in the Dirac semimetal and 3D topological insulator state, respectively. We have observed for both states a negative MR with current either parallel or transverse to the in-plane magnetic field, but with a different dependence of MR on $\vec{B}$ strength. Our results are inconsistent with the chiral anomaly and suggest that other mechanisms may be responsible for negative MR in the Dirac/Weyl semimetal phase of \aSn{}.  We also discuss several factors in sample design and material quality that may be contributing to the incongruous observations of MR reported in studies of strained \atn{} films 
\end{abstract}

\pacs{75.47.-m, 03.65.Vf, 72.20.My }

\maketitle
\doi{xxxxxxxxxxxx}

\section{Introduction}

 The $\alpha$ allotrope of tin (\atn{}, gray tin) is a  zero-gap semiconductor with an inversion symmetric diamond lattice and an s-p band inversion \cite{Groves:63,Pollak:70,Huang:17}. The conduction band minimum and valence band maximum are degenerate at the  $\Gamma$ point. Band degeneracies of this type can be tuned by strain by reducing the crystal symmetry thereby removing the degeneracy and changing the band structure \cite{Roman:72}. A compressive uniaxial strain lifts the $\Gamma$-point degeneracy and opens a positive energy gap between the $\Gamma_8^+$ branches that results in a 3D topological insulator (3D TI) phase \cite{Fu:07}. Oppositely, a tensile  uniaxial strain 
 induces an overlap between the inverted light-holes (invLH) and the heavy-holes (HH) bands of the unstrained material,
 \figintro 
 which creates  a pair of band singularity points at $k=(0,0,\pm k_D)$, characteristic of a Dirac semimetal (DSM) \cite{Huang:17} (Fig. \ref{figintro} a). If the time-reversal symmetry is further removed by applying an external magnetic field, a Weyl semimetal (WSM) phase is expected to emerge where the Dirac points split into chiral, spin-resolved pairs. WSM materials have been of considerable interest in the past decade as electrical conductors with nontrivial properties derived from topology in k-space \cite{Volovik:07}. 
 
 Several recent studies have focused on strained \atn{} as a simple, elemental platform for studies of transport phenomena in WSMs. Epitaxial films of \atn{} are usually grown on CdTe or InSb substrates (or buffer layers) that have slightly smaller lattice constants than \atn{}, leading to a compressive, biaxial in-plane strain and a corresponding {\em tensile} strain along the axis normal to the film due to the Poisson effect, thereby producing the DSM state. In a magnetic field (i.e. in the WSM state), the  analog  of the Adler-Bell-Jackiw chiral anomaly \cite{Adler:69,Bell:69} is expected to manifest a strong enhancement of conductance with magnetic field,  i.e. a negative  magnetoresistance (MR), when the electric current flows along the direction of the magnetic field\cite{Burkov:14,Son:13}.  The suppression of the longitudinal MR but not the transverse MR is, therefore, expected to occur.

Several magnetotransport studies of \atn{} films have recently attributed negative  MR to the chiral anomaly  \cite{Polacz:24,Alam:24,Basnet:24, Li:25}.The origin of the negative MR  and the role of chiral anomaly-related effects in transport in strained \atn{} films remains challenging to understand. The available transport studies in \atn{} films do not yet present a consistent picture between different growth techniques and substrate choices. In addition, we note that negative MR has been observed in unstrained bulk crystals of \atn{} \cite{Tufte:61,Hinkley:64,Huang:17}. Experiments on mechanically strained, bulk {H}g{T}e \cite{Takita:79}, which has a very similar band structure to \atn{}, show a pronounced negative MR  under a compressive uniaxial strain, when Dirac points are absent and there is instead a strain-dependent gap between the conduction and the valence bands as in a 3D TI.

The strain  of the \atn{} film, therefore, determines whether the film is a topological insulator or a DSM. To test the hypothesis that the chiral anomaly is responsible for negative MR in the WSM phase of \aSn{}, we reverse the strain to achieve a 3D TI state to remove a necessary condition for the chiral anomaly.  Since there is no available substrate with a larger lattice constant than \atn{} to induce the desired strain, we instead alloy \atn{} with Ge to shrink its lattice constant and reverse the mismatch strain to change the material from a DSM to  a 3D TI.  The relevant band structure changes due to alloying are negligible for a low Ge composition, thereby maintaining the topological properties of pure \atn{}.  \cite{Polak:17} We find negative  longitudinal  MR  in both the 3D TI state and the  WSM state, which casts doubt on the chiral anomaly as the  dominant mechanism for negative  MR in \atn{}, and reach a similar conclusion from the analysis of the transverse MR. 

Because material quality and sample preparation can significantly affect magnetotransport (see Appendix, section \ref{subsec:review}), great care was taken in preparing samples in this study. In contrast to films grown on Ar$^+$ ion-bombarded InSb substrates or plastically relaxed CdTe buffer layers, the $\alpha$-Sn and $\alpha$-SnGe films of this study were grown by molecular-beam epitaxy directly on commercially available CdTe substrates with no additional epilayers between the substrate and the films. Unlike InSb,  CdTe has very low conductivity even at room temperature, and, therefore, does not contribute stray  current paths in   magnetotransport measurements. Because the films are monocrystalline with negligible levels of crystallographic defects, we expect the observed magnetotransport behavior to be attributed solely to the intrinsic $\alpha$-Sn properties.

\section{Experiment}
\subsection{Material synthesis and characterization}
CdTe(001) substrates (1 cm x 1 cm) from JX Advanced Metals Corporation were degreased and then etched for 30 seconds in a solution of 0.5\% by-volume bromine in methanol. The substrates were then rinsed in methanol and de-ionized water and immediately loaded into vacuum. Within a custom MBE chamber with a base pressure of  $3\times  10^{-8}$ Torr, the substrate was slowly heated to 140-145 $^\circ$C to desorb the amorphous Te layer left behind by bromine etch. \cite{Haring:83} In situ monitoring of the surface by reflection high-energy electron diffraction (RHEED) indicates Te removal by the appearance of the CdTe (2$\times$1) surface reconstruction diffraction pattern. \cite{Benson:86} The substrate  was then radiatively cooled  until its temperature was below 20 °C, as measured by the thermocouple on the substrate heater. Sn and Ge were co-deposited from effusion cells at a combined rate of 1.5 \r{A}/s while the surface is monitored by RHEED during deposition.  An alloy composition of 2\% Ge was targeted to ensure biaxial tensile strain while at the same time keeping total strain energy below thresholds for dislocation glide and phase transformation in a desired thickness range of 150-200 nm. To determine film thickness, strain, composition and crystallinity, high-resolution X-ray diffraction (XRD) was performed as described in the Appendix, section \ref{subsect:struct}. As shown in Table \ref{table1}, the pure  \aSn{} film was biaxially compressive strained, and the \aSnGe{} samples were biaxially tensile strained.
\ttable

\subsection{Device Fabrication}
We used a photolithographic process  specifically modified  to be compatible with the  low phase transition temperature of epitaxially stabilized $\alpha$-Sn. This approach eliminates errors due to sample geometry that are inherently present  in the simpler, process-free van der Pauw method   \cite{Marion:82}, or in Hall bars defined by crude film scratching.  Hall bar samples were prepared as defined by ASTM specifications for a ``Bridge-type" specimen. \cite{ASTM:16}.  The samples were mounted on a custom circuit board and electrically connected to the transport circuit with 25 $\mu$m gold wires attached to the sample with diluted silver paint (Leitsilber 200, Ted Pella, Inc). The resulting low-resistance connections to \atn{}  were robust over multiple cool-downs and did not show any significant change in resistance with time. CdTe substrates used for film growth showed infinite resistivity in room temperature tests. Details of the sample fabrication protocol are discussed  in the Appendix,  section \ref{subsec:fab}.

\subsection{Magnetotransport}
The coordinate system for the  sample  is shown in Fig. \ref{figintro}b.  The long axis of the Hall bar is parallel to the $x$-axis and is the direction of the current, $\vec{I}=I\hat{x}$.  The magnetic field, $\vec{B}$, has the components 
\begin{equation}
   \vec{B}=B_\parallel \hat{x} + B_\perp \hat{y}+ B_z \hat{z}
\end{equation}
where $B_\parallel$ is the longitudinal field component parallel to the current, $B_\perp$ is the field in the film plane transverse to the current, and $B_z$ is the  field component orthogonal to the sample plane. 
The electric  current entering the sample via $C_1$ was recorded with a Keithley 2400 source meter. The current exiting through $C_2$ was also recorded to ensure proper isolation of the transport circuit from the cryostat during measurements. The longitudinal voltage drop $V_{xx}$ was measured across the $B_1$ - $B_2$ terminals with the DL Instruments 1201 low noise amplifier or the Agilent 34401A multimeter. The current path shown in the figure was used for samples $S2$ and $S3$. For the  $S1$ sample, the contacts $B_2$ and $B_3$ were used to establish the current path because  a film crack  occured during cooldown that disabled the contact  $C_1$. The pair $A_2$-$A_3$ was used for the $V_{xx}$ measurements in sample $S1$.  The magnetoresistance MR is defined as 
\begin{equation}
\mbox{MR}= 100 \% \times \frac{R(\vec{B})-R(\vec{B}=0)}{R(\vec{B}=0)}
\end{equation}
All MR data were recorded  at temperature $T$= 5 Kelvin in a probe with sample in vacuum, in a liquid helium cryostat equipped with a 9-Tesla magnet (American Magnetics, Incs solenoid magnet \#13665, Cryogenics Limited power supply model SMS120C.)

\figa{}
The results of  MR measurements are presented in Figures \ref{figa} and \ref{figcdcombo}. First,  we note the huge difference in the strength of the MR observed with the  in-plane and out-of-plane magnetic field direction. Figure  \ref{figa} shows MR in sample $S1$ (\atn{}) with two magnetic field orientations:  $\vec{B}=B_z\hat{z}$, orthogonal to the film plane,  and  $\vec{B}=B_{||}\hat{x}$, parallel to the current direction. With  $\vec{B}$ normal to the plane,  the film resistance   increases approximately 4-5 times  between zero and 8 T. Under $\vec{B}=B_{||}\hat{x}$, the MR decreases, by about 30\%, in the same field range.  \aSnGe{} films showed comparable increases in MR under $\vec{B}=B_z\hat{z}$ (Appendix, section \ref{subsec:Hallfield}). The rest of the present discussion focuses on the MR features under in-plane magnetic field.

Figure \ref{figcdcombo} shows MR as function of the magnetic field for two in-plane orientations: parallel to the current,  $\vec{B}=B_{||}\hat{x}$ (Fig. \ref{figcdcombo} a, b), and  orthogonal to the current, $\vec{B}=B_\perp \hat{y}$, (Fig. \ref{figcdcombo} c, d). 
The MR of the pure \atn{} sample decreases nearly monotonically with field when  $\vec{B}=B_{||} \hat{x}$, and  reaches  $\approx 30$ \% at 8 T. A very small, less than 0.2 \%, positive MR  is observed at fields below 0.5 T (Fig. \ref{figcdcombo}, b). When the in-plane magnetic field is normal to the current, $\vec{B}=B_\perp \hat{y}$, we also observe negative MR as the magnetic field increases. The film resistance decreases even  faster with $B_\perp$ than with $B_{||}$ up to 3 T. The MR remains negative until the field reaches approximately 5.5 T and becomes positive at higher fields.

Surprisingly, the effects observed in samples made from \aSnGe{} films are qualitatively similar. We have studied two samples containing Ge, $S2$ and $S3$, with slightly different thicknesses of 155 nm and 206 nm, respectively. The data show that adding Ge somewhat reduces the effect of the magnetic field on MR compared to the pure  \atn{} film. However, both \aSnGe{} samples still show  negative MR under $B_{||}$ in the entire field range (Fig. \ref{figcdcombo}, a, b). With the in-plane field  transverse to the current, the MR($B$) dependence is non-monotonic in both  \atn{} and \aSnGe{}  films, and the MR remains negative in the samples containing Ge across the entire range of available fields. 

\section{Discussion}
 \atn{} and $\alpha$-SnGe samples studied in this work are in  opposite strain states. As a result, while \atn{} has Dirac/Weyl bulk points in the band structure, $\alpha$-SnGe is a 3D topological insulator. The observation of negative  MR   both in  the \atn{} and the $\alpha$-SnGe sample is  therefore incompatible with the interpretation of chiral-anomalous effects as the origin of negative MR. Additionally,   the  chiral current  is proportional to $\vec{E}\cdot \vec{B}$, therefore, negative MR due to the chiral anomaly  cannot be induced by  $B_{\perp}$. Yet, our measurements show negative MR not only when the field is parallel to the current, but also when the field is orthogonal to the current flow, both in \atn and $\alpha$-SnGe.

Polaczy\`{n}ski et al. \cite{Polacz:24} studied \atn{} films grown on a GaAs substrate capped with a 4$\mu$-thick CdTe buffer layer. They  observed a negative  20-30 \% MR with $\vec{B}||\vec{I}$ at magnetic fields 10-15 T in films ranging from 50 nm to 200 nm thick. At low fields, they found a positive  MR of order of several percent at fields 1-2 T (Fig. 6 in \cite{Polacz:24})  which turned negative as the magnetic field increased.  Interestingly, the effect was less strong in the 150 nm sample that was  closest to our samples in thickness. The data in \cite{Polacz:24} also indicate that the strength  of the negative MR effect is thickness and temperature dependent.  In the 50 nm  sample, the MR(B) is roughly the same between 0.3K and 77 K, however, in the 200 nm sample, the MR is present at 77 K  in a wide range of magnetic fields but remains roughly zero at 0.3 K and  the negative MR develops only at  fields above 7-8 T, with the low-B data dominated by Shubnikov-de Haas oscillations. In our experiment, we observe virtually no positive MR even at the smallest fields and, strikingly, the negative MR is present in both $\vec{B} || \vec{I}$  and $\vec{B} \perp \vec{I}$ field orientations. Alam et al.  \cite{Alam:24} found a decrease in resistance similar to \cite{Polacz:24} on comparable substrates, and  showed that the negative MR  disappears when the magnetic field is tilted out of plane by a small angle.
\figcdcombo{}

Basnet et al. \cite{Basnet:24} investigated MR both  in pure \atn{} and $\alpha$-SnGe films. They also observed a much stronger MR at low fields $B<2.5$ T but find {\em positive} MR in the entire range of magnetic fields, clearly inconsistent with our measurements.  Li et al. \cite{Li:25} have studied \atn{} grown on CdTe substrates capped with {InSb} and doped with phosphorus in different concentrations. The authors investigated anisotropy of MR with respect to the in-plane magnetic field orientation (Fig. 2b in \cite{Li:25}). They find that the magnetic field dependence of MR  changes from monotonically increasing with $B$ when $\vec{B}\perp \vec{I}$ to non-monotonic, with MR increasing up to $\sim$ 4-5 T and then decreasing as the field increases, under $\vec{B} || \vec{I}$. Interestingly, the MR itself, unlike the differential slope $d(\mbox{MR})/dB$ reported in \cite{Li:25}, remains positive for all field orientations. These observations differ from ours, because we observe negative MR in a wide range of magnetic fields, for  both $B_{\perp}\hat{y}$ and $B_{||}\hat{x}$ field orientations.  It may be possible that these incongruous observations are due to various extrinsic effects caused by sample design as outlined in the Appendix (section \ref{subsec:review}).  Samples in this study were prepared to avoid confounding factors.

It is instructive to compare the magnetoresistance  of strained \atn{} films to bulk measurements in  {H}g{T}e.   Takita et al. \cite{Takita:74,Takita:79} studied  longitudinal \cite{Takita:79} and transverse \cite{Takita:74} MR in  3D HgTe crystals under uniaxial {\em compression}.  The compressive uniaxial strain opens the gap between the degenerate valence and the conduction band, that the authors confirmed by temperature-dependent transport measurements \cite{Takita:79} .  The material strain was, therefore,  opposite  of the conditions in experiments on \atn{} films, including ours, where a uniaxial {\em dilation} normal to the film plane arises from the biaxial compression in the film plane due to the substrate lattice mismatch.  Therefore, the samples in \cite{Takita:74,Takita:79}  were  in the  topological insulator state, with no overlap of LH and HH bands and no corresponding Dirac nodes. Moreover, there was no  crossing of the Landau levels originated from LH and HH bulk bands in a wide range of  magnetic fields \cite{Yoshizaki:76}. Yet,  the authors observed a dramatic negative longitudinal ($\vec{B} \,|| \,\vec{I} $) MR   that became stronger as the compressive force increased \cite{Takita:79}. It should be noted that the electric current direction in this experiment was along the axis of the uniaxial strain and therefore the experiment is not entirely equivalent to transport measurements in biaxially strained films, where the current is in the plane and is orthogonal to the strain axis. Importantly, the negative MR in this experiment occurs in the gapped, topological insulator,  phase,  where the Dirac points are absent. This suggests that the mechanisms responsible for the negative MR are not related to the  chiral anomaly. Negative {\it transverse}  MR ($\vec{B} \perp \vec{I}$)  was also observed \cite{Takita:74}, with magnetic fields  ranging from  zero up to several Tesla \cite{Takita:74}, and increased with strain.   The authors \cite{Takita:79,Takita:74} attributed these observations to changes in  the separation between Landau levels originating from LH and HH bulk bands, which leads to a reduction of the strain-induced gap with increasing magnetic field in inverted band structure materials. \cite{Yoshizaki:76}.

Negative MR in  HgTe and \atn{}  may  occur due to  the  mechanism proposed in Refs. \cite{Khaetskii:84,Dyakonov:17}. The effect arises from the strong spin-orbit interaction in these materials. Carrier scattering by impurities in presence of spin-orbit coupling results in the  coupling of the  electrical and spin currents. In the absence of the magnetic field, the mutual transformation of the electrical and spin currents {\it reduces} the apparent electrical conductivity \cite{Dyakonov:17}. An external magnetic field acts on both the orbital and the spin degrees of freedom, and destroys the spin current  by changing the relative orientation of the spins and the velocity of carriers, which increases the conductivity, thus  leading to a negative MR. In this scenario, the effect occurs at  both transverse and longitudinal magnetic field with respect to the current flow, in agreement with our measurements. This is expected to occur for magnetic fields ranging from zero to  $\omega_c \tau_{tr}\equiv \mu_{tr} B \gtrsim 1$, or  between 0 and $\sim$ 1.5-1.8 T ( $\mu_{tr} \simeq 6000 cm^2/V s $ ) in our experiment. We also note that  Takita et al.  observed a negative longitudinal magnetoresistance in  bulk HgTe samples even
at zero strain  (Ref. \cite{Takita:79}, Fig. 5a.),  consistent with the spin-orbit-related mechanism proposed in Ref. \cite{Khaetskii:84}. 

\section{Conclusion}
It is apparent that the manifestation of chiral anomaly in transport in \atn{} remains insufficiently understood, as the data  from available magnetoresistance studies do not yet form a consistent picture. While some groups find negative magnetoresistance under in-plane field, positive MR is also sometimes observed. Our observation of negative MR even with $\vec{B}\perp \vec{I}$ indicates that there are mechanisms other than the chiral anomaly that lead to a negative MR in the strained-induced DSM phase of  \atn{}. Also, experiments in strained HgTe, where the band structure is very similar to \atn, show that negative MR can be induced by strain in the topological insulator phase, i.e.  when the positive sign of the uniaxial strain is {\em opposite} of its value in epitaxial strained films. We suggest that the conflicting observations reported in the literature on transport in strained $\alpha$-Sn films may be due to extrinsic factors resulting from  differences in  sample design and preparation. We therefore hope that our results will stimulate further studies of the origin of negative magnetoresistance in  Dirac and Weyl semimetals. 
\section{Acknowledgments}
This work was supported in part by the Air Force Office of Scientific
Research (Grant No. FA9550-AFOSR-23RYCOR05) and by University of Cincinnati.
\section{Appendix}
\subsection{Sample-specific considerations in magneto-transport studies of \atn{} } \label{subsec:review}

Material quality is fundamentally important when investigating intrinsic magnetotransport properties, since material defects and sample inhomogeneity often lead to spurious results \cite{Pippard:89}.  Several different methods have been used to prepare $\alpha$-Sn and $\alpha$-Sn$_{1-x}$Ge$_x$ samples for electrical measurements.  Films prepared in different ways do not  behave the same.  Some recent reports point to defects, morphology, and impurities in $\alpha$-Sn films as important factors in electronic transport properties. Given the range of observations in the literature and their interpretations, we review some of the sample preparation methods and resulting material characteristics to highlight these issues and provide context for evaluating the various reported results.

Working with $\alpha$-Sn is challenging due to its low phase-transformation temperature to metallic $\beta$-Sn at 13°C for bulk material.  Before 1981, monocrystalline $\alpha$-Sn samples were obtained by precipitation from a mercury amalgam or by carefully cooling $\beta$-Sn whiskers until they converted to $\alpha$-Sn \cite{Busch:60}. These samples required constant refrigeration to avoid phase transformation and the material was extremely brittle. However, magnetoresistance and other measurements were made on bulk samples of $\alpha$-Sn.  \cite{Ewald:68,Lavine:71} In 1981, Farrow et. al. demonstrated growth of $\alpha$-Sn as an epitaxial film stable over 70°C on nearly lattice-matched InSb and CdTe substrates, allowing experiments to be done at a more convenient room temperature. \cite{Farrow:81} Since then, nearly all published studies of $\alpha$-Sn have been carried out in the epitaxial-film format. 

While heteroepitaxy greatly relaxes temperature constraints for phase stability, it complicates characterization of $\alpha$-Sn’s band structure and electrical properties.  First, $\alpha$-Sn (6.489 \r{A})  \cite{Nbs:53} is not perfectly lattice-matched to available substrates, causing it to be biaxially and compressively strained when grown on InSb (6.4793Å) \cite{Straumanis:65} and CdTe (6.481 Å) \cite{Nbs:64}, the most closely lattice-matched substrates.  The mismatch strain alters the electronic band structure of $\alpha$-Sn that is sensitive to small strain due to its gapless nature. \cite{Cardona:67,Roman:72,Huang:17} The film's total strain energy increases as the thickness increases, eventually leading to crystallographic defects and phase transformation.\cite{Matthews:74,Zunger:89,Reno:89,Song:19,Edirisinghe:25} Dislocations cause electronic defect-states and inhomogeneous strain within the crystal that can locally affect band-energies.  Furthermore, the properties of very thin films ($<$50-nm thick) are modified by quantum effects.\cite{deCoster:18,Khaetskii:24} Thus, an epitaxial film of $\alpha$-Sn may not behave as a homogeneous, bulk material often assumed in practice.

The $\alpha$-Sn film’s microstructure and resulting properties can depend significantly on the preparation of the substrate’s surface prior to film growth.  The substrate’s surface oxide must first be removed to expose the bulk crystal structure, or $\beta$-Sn will nucleate instead.  Oxide can be removed in a few different ways depending on the substrate. The most common method to clean substrate surfaces is to bombard them with Ar$^+$ ions followed by thermal annealing to heal crystallographic damage.  This technique has been used since the inception of MBE growth of $\alpha$-Sn.  \cite{Farrow:81} While it may remove surface oxides and contaminants and produce atomic ordering of the surface as evidenced by in situ electron diffraction, the subsurface region can retain crystallographic damage affecting the electrical properties of the substrate as revealed in electrical measurements.  Studies of InSb cleaned by ion bombardment and annealing showed that the surface of a p-type InSb substrate becomes n-type, manifesting a two-dimensional electron gas (2DEG).  \cite{Yuen:89,Yuen:91,Yuen:93,Liu:95,Zheng:86}   An additional charge layer with high-mobility carriers at the interface can confound the magnetoresistance in an $\alpha$-Sn film. Additionally, the valence band offset of InSb with $\alpha$-Sn is approximately 0.4 to 0.6 eV, \cite{John:89,Ekpunobi:99,Qteish:92,Lambrecht:90} forming a Type-III heterojunction; the $\Gamma_8$ - $\Gamma_6$ band gap of InSb (~0.17eV) is smaller than the corresponding inverted bandgap $\Gamma_7^-$ - $\Gamma_8^+$ of $\alpha$-Sn  (-0.41eV) \cite{Groves:70,Carrasco:19}, creating an unusual electrical interface that has not been theoretically explored.

Crystallographic damage caused by ion bombardment can also propagate from the substrate into the epitaxial film.  In a recent report \cite{Madarevic:20}, $\alpha$-Sn films grown on ion-bombarded and annealed InSb(001) are described as having grain sizes of $\sim$15 nm for films roughly the same thickness.   Similarly, the authors of \cite{Madarevic:22} used Ar$^+$ sputtering to prepare their InSb substrates and characterized their 30-nm-thick films as having a “granular morphology” with grain sizes of 10 to 20 nm. These authors concluded that grain-boundary scattering reduced the mobility of bulk carriers to make Dirac charge carriers more “accessible” to measurement.  A completely different approach to growing $\alpha$-Sn with a similar morphology is by DC magnetron sputtering of Sn onto a Si(111) substrate that first undergoes several sputtering and annealing cycles to remove the surface oxide. \cite{Ding:21} The Sn films are deposited at room temperature without subsequent annealing.  Films that are $<$7 nm thick are predominantly $\alpha$-Sn while films $>$10 nm thick are $\beta$-Sn.  The lateral grain size is 15 to 20 nm, like the films in \cite{Madarevic:20}.  These authors also claim that the topological surface states dominate in these films because the grain boundaries trap bulk carriers while leaving surface states unaffected.  These observations imply that polycrystalline films of $\alpha$-Sn have qualitatively different properties than monocrystalline films.

In addition to inducing crystallographic defects in the film, the substrate can couple into the electrical measurements of the film.  InSb itself has a very high mobility and magnetoresistance \cite{Hu:08}, so significant current may be shunted through the substrate during a transport measurement of an epitaxial $\alpha$-Sn film, causing the InSb properties to confound the measurement. \cite{Massetti:25,Shiomi:16}   Furthermore, it has also been observed that atoms from the InSb substrate’s selvedge may diffuse into the $\alpha$-Sn film during growth, changing the film’s majority carrier type and concentration.  \cite{Engel:24}  The termination of the bare InSb surface in vacuo is sensitive to temperature and Sb-overpressure, resulting in either In-rich or Sb-rich surfaces.  The $\alpha$-Sn film can incorporate these surface atoms inhomogeneously, depending on the Sn deposition rate and substrate temperature. \cite{John:89,Betti:02,Magnano:02,Chen:22} Given these various confounding factors, any electrical transport measurements of $\alpha$-Sn on InSb should be considered with great care.

To avoid some of these complications, growth of films directly on CdTe substrates \cite{Reno:89,Carrasco:19,Bowman:90,Gomez:03,Tu:89,Vail:20} and Cd$_{1-x}$Zn$_x$Te substrates \cite{Kiefer:17} is also possible.  CdTe has a bandgap of $\sim$1.5eV at 300 K and wafers can have a resistivity of $10^{10} ~\Omega\cdot$cm at 300K, so electrical shunting through the substrate should be negligible, though unintentional doping of the film with Cd and/or Te from the substrate during growth may still be possible. Besides using ion-bombardment and thermal annealing, the surface oxide of CdTe substrates can also be removed by etching in a bromine-methanol solution to avoid crystallographic damage.  This wet etching leaves an amorphous layer of Te on the surface that requires thermal desorption in vacuo prior to deposition of $\alpha$-Sn. \cite{Haring:83}  The in situ monitoring of the surface by RHEED indicates Te removal by the appearance of the CdTe (2×1) surface reconstruction diffraction pattern.  \cite{Benson:86} This substrate preparation process minimizes crystallographic defects, strain inhomogeneity, and substrate-related doping within the film.

Some effort has been made to circumvent the need for CdTe substrates for electrical isolation by using CdTe buffer layers on various substrates.  Recently, a couple of magnetotransport studies of $\alpha$-Sn films on CdTe(001) buffer layers were published. The authors of Ref.  \cite{Polacz:24} state that the key to their successful characterization of their films is the hybrid substrate of a 4-µm film of CdTe on semi-insulating GaAs (001).  A CdTe buffer layer, however, necessarily has defects from plastic relaxation of strain when grown on the mismatched GaAs substrate.  The buffer layer was reported to have an XRD rocking curve FWHM of ~300 arcsec, about ten times greater than typical CdTe substrates.  It is likely the $\alpha$-Sn films grown on these hybrid substrates inherited similar levels of defects from the buffer layer as well as inhomogeneous strain, perhaps placing the films in the same category as those grown on sputter-cleaned surfaces.  In comparison, the authors of \cite{Basnet:24} used a CdTe buffer layer grown on InSb (001) which is nearly lattice-matched.  The magnetotransport behavior of $\alpha$-Sn films in  \cite{Polacz:24} and  \cite{Basnet:24} disagree both qualitatively and quantitatively, indicating that the use of these CdTe hybrid substrates may not overcome the several problems previously mentioned.

The possible presence of $\beta$-Sn islands within the $\alpha$-Sn film poses yet another problem, especially at low-temperatures ($\le$4K) where many magnetotransport measurements are made.  $\beta$-Sn islands embedded in epitaxial $\alpha$-Sn films grown on InSb (001) have been observed since the earliest epitaxial films by Farrow et al., who emphasized that insufficient annealing of ion-bombarded substrates led to high densities of $\beta$-Sn islands.  \cite{Farrow:81}   A recent experiment focused on the role of $\beta$-Sn islands in the superconducting transition (SC) temperatures of $\alpha$-Sn films grown on two different substrates: an InSb(001) substrate with an InSb(001)-c(2$\times$4) buffer layer, or a GaAs(001) substrate with a GaSb/InSb(001)-c(2$\times$4) buffer layer.  \cite{Ding:23}   The film grown directly on InSb(001) had no SC transition, while the films grown on the composite substrate had a sharp SC transition at 3.7K, near the bulk value for $\beta$-Sn, with two additional resistance drops at slightly higher temperatures.  The authors attributed the SC behavior to $\beta$-Sn islands in the films grown on the composite substrates due to increased crystallographic defects caused by plastic relaxation in the buffer layer.  XRD diffraction clearly identified the $\beta$-Sn islands in the as-grown films on the composite substrate while it could not detect any $\beta$-Sn in the films grown directly on InSb(001).  However, as the film on the InSb(001) aged by 10 days, slight resistance drops were observed at the same temperatures as the films grown on the composite substrate, until a full SC transition occurred in the film after two months of aging, indicative of the increasing amount of $\beta$-Sn present in the $\alpha$-Sn film with time.  This experiment demonstrated the probable formation of $\beta$-Sn in films grown on relaxed buffer layers and even the insidious formation of $\beta$-Sn within films grown directly on a nearly lattice-matched substrate.  Many researchers do not report on the presence or absence of $\beta$-Sn, likely because the prominent diffraction peaks for $\beta$-Sn (2$\theta$ $\approx$30°-32°) are far away from those of $\alpha$-Sn(004) (2$\theta$ = 56.7°) such that $\beta$-Sn is unnoticed unless intentionally sought.  It should also be noted that a low density of $\beta$-Sn islands will also evade detection by RHEED during growth.  $\beta$-Sn is 26\% denser than \atn{} and takes up less volume for the same mass.  The islands, which start at the substrate surface, have tops that sit below the surface of  the \atn{} film and are not visible by the grazing-incident electron beam.

In summary, not all $\alpha$-Sn films reported in the literature are qualitatively the same.  Samples vary by substrate, thickness, strain, impurities, defects and grain size.  All these characteristics are not usually evaluated or well-controlled in each experiment.  Given the variety of samples reported and the presence of a substrate that may be coupled into the electrical measurements, it is difficult to conclusively determine the intrinsic properties of the films in any given experiment or even collectively.  The surface damage in InSb substrates prepared by ion-bombardment can create both electrical changes and structural damage within the substrate, confounding magnetoresistance measurements.  Growth on sputter-cleaned substrates or relaxed buffer layers may produce grain boundaries in the film that have been claimed in recent experiments to depress bulk-carrier mobility so Dirac-fermion charge-carriers in topological surface states can dominate transport.  Surface preparation methods, buffer layers, sample thickness, and mismatch strain can also promote growth of $\beta$-Sn islands in the film, further confounding the measurement.  Thus, all aspects of sample preparation must be carefully considered, characterized, and controlled to avoid a range of interfering factors in magnetotransport measurements.

\subsection{Fabrication and characterization of samples}\label{sect:fabcharsamples}
\subsubsection{Structure}\label{subsect:struct}
\figxrdd{}

Structure determination was performed by using x-ray diffraction with a PANalytical Empyrean X-ray diffractometer configured with a CuK$\alpha_1$ radiation source and a triple-axis analyzer. XRD $2\theta -\omega$ scans for all films showed well-defined (004) film peaks with Pendellösung fringes.  No peaks for $\beta$-Sn were detected immediately after growth.  The composition and thickness of the films were determined by fitting a dynamical diffraction model to the data using the commercial software X’Pert Epitaxy. A representative dataset and model are shown in Fig.  \ref{figxrdd}. Reciprocal space maps (RSMs) of the (004), (335), ($\overline{33}$5) (3$\overline{3}$5) and ($\overline{3}$35) reflections were recorded to determine the in-plane and out-of-plane lattice parameters of the substrates and films.  Multiple RSMs were used to account for substrate miscut and tilt.  Representative maps for a (004) and (335) reflections are shown in Fig. 
\ref{figxrd}.  The RSMs indicate that the films have the same in-plane lattice constants as the substrate, and we conclude that the films are elastically strained with negligible relaxation. The lattice parameters of the films are derived from the peak positions using Bragg’s law, and the epitaxial strain in the film was determined using standard elastic relationships for cubic materials. Results are shown in Table I.  
\figxrd{}

\subsubsection{Device Fabrication}\label{subsec:fab}
Standard photolithography methods require temperatures above the phase transition point for epitaxially stabilized \atn{}, so lower temperatures were used at longer times to achieve adequate results.  We start by forming an etch mask by coating the sample with a 500-nm thick layer of photoresist (S1805, Kayaku Advanced Materials), soft-baked at 75\textdegree C for 15 minutes on a hotplate at atmospheric pressure with hexamethyldisilizane (HMDS, Sigma Aldrich) as an adhesion promoter.  After patterning the etch mask by exposing to UV light and developing with Microposit 300 MIF for 30 seconds, we etched the exposed film using a Cl-based reactive ion etch process.  We then removed the photoresist by first treating the etched sample for 4 minutes with O$_2$ plasma in a barrel asher followed by solvent cleaning. We patterned the Ti/Au metal contacts and traces using a liftoff process.  A finished Hall bar device is shown in Figure \ref{figdevdim}. Hall bar channels were nominally 3000 µm long (L$_c)$ and 600 µm wide (w$_c$). The voltage probes were 600 µm long (L$_p$), 120 µm wide (w$_p$), and spaced 660 µm apart (d$_p$). 
\figdevdim{}  
\subsubsection{Transport}\label{subsec:Hallfield}
\figmrbz{}  

Magnetotransport with out-of-plane magnetic field ($\vec{B} = B_z \hat{z}$) is presented in Fig. \ref{mrbz}. The measurements on the pure film are shown in black, and the Ge-alloyed films with increasing thickness are shown in orange and green, respectively.  All samples show qualitatively the same parabolic dependence on $\vec{B} = B_z \hat{z}$, with MR at maximum field being $\approx$ 4 times larger than zero-field values. Alloying the samples with Ge slightly reduces the field effect and in-plane MR data follows the same trend with Ge-alloying. 

\fighallge{} 
Hall measurements with the current flowing along the length of the Hall bar, voltage probes transverse to the current flow, and magnetic field normal to the film plane were used to estimate carrier density (Fig. \ref{LinearGeFit}). For a single-band carrier  model, the Hall resistivity is
\begin{equation}
    \rho_{xy} = t\,\frac{V_{xy}}{I} = \frac{B_z}{n q}
    \label{Hall_density}
\end{equation}
where $\rho_{xy} = t\,\frac{V_{xy}}{I}$ is the transverse resistivity, $V_{xy}= V_{A^i} - V_{B^i}$ is the transverse voltage, defined as the voltage difference between pairs of contact $\{A_i, B_i\}$ as shown in Fig. \ref{figintro}b, $I$ is the current through the sample, $t$ is the sample's thickness, $B_z$ is the magnitude of the out of plane magnetic field, q is the carrier charge, and n is the carrier density. The slope $\frac{d\rho_{xy}}{dB}$ is inversely proportional to $n$ and the sign of the slope is the sign of the carrier.

The single-band resistivity at zero magnetic field,  $\rho_{xx}$ can be written as

\begin{equation}
    \rho_{xx} = \frac{1}{n |q| \mu}
\end{equation}
where $\mu$ is the mobility of the sample. The sample's mobility can be determined if $\rho_{xx}$ and $n$ are known. From Fig. \ref{LinearGeFit}, the slope of the linear fit gives the apparent carrier density of the Ge-alloyed samples and the carrier sign (n-type). Combined with the resistivity values from Table \ref{table1}, this gives the carrier densities  $\sim 10^{17} \ cm^{-3}$ and mobilities are $\sim 6000 \ \frac{cm^2}{V \cdot s}$. The carrier densities and transport mobilities obtained from the fits are given in  Table~\ref{TransportTable}. 
\tttable{}

\newpage
%
\end{document}